# Pulsatile Drug Delivery System Based on Electrohydrodynamic Method

Yi Zheng*, Yuan Zhang, Junqiang Hu and Wenle Gao
Department of Mechanical Engineering, Columbia University, New York, NY 10027
* Email: yz2308@columbia.edu

**Abstract**

Electrohydrodynamic (EHD) generation, a commonly used method in BioMEMS, plays a significant role in the pulsatile drug delivery system for a decade. In this paper, an EHD based drug delivery system is well designed, which can be used to generate a single drug droplet as small as 2.83 nL in 8.5 ms with a total device of $2 \times 2 \times 3$ mm$^3$, and an external supplied voltage of 1500 V. Theoretically, we derive the expressions for the size and the formation time of a droplet generated by EHD method, while taking into account the drug supply rate, properties of liquid, gap between two electrodes, nozzle size, and charged droplet neutralization. This work proves a repeatable, stable and controllable droplet generation and delivery system based on EHD method experimentally as well as theoretically.

**Nomenclature**

| | | |
|---|---|---|
| $E$, | [V/m]: | Electric field strength |
| $\dot{m}$, | [kg/min]: | Drug supply rate |
| $\rho_d$, | [kg/m$^3$]: | Drug density |
| $\varepsilon$, | [F/m]: | Drug fluid permittivity |
| $\mu$, | [Pa·s]: | Drug fluid viscosity |
| $a$, | [m]: | Radius of nozzle |
| $h$, | [m]: | Height of liquid droplet |
| $d$, | [m]: | Gap between tip of nozzle and ground electrode |
| $\gamma$, | [N/m]: | Surface tension |
| $\theta_c$, | [rad]: | Contact angle |
| $\rho_e$, | [C/m$^3$]: | Spatial charge density |
| $r$, | [m]: | Radius of droplet |
| $r_{eff}$, | [m]: | Effective radius of droplet |
| $F_e$, | [N]: | Electric force |
| $F_s$, | [N]: | Surface force |
| $\zeta$, | [V]: | Surface electric potential |

## I. Introduction

Microfluidics and related nano/micro-technology have been active research fields for several years. Numerous microflow devices such as microchannels [1,2], flow sensors [1,16,17], microvalves [3], micropumps [2,5,9,10-12], electrowetting [4,6], electro-hydrodynamics (EHD) [13-15,21-24] and magnetic-hydrodynamic [7] have been reported. A droplet dispenser using the microfluidic chip format [8] has recently found increasing popularity due to its potential for massively parallel array ability and scalability of its size [18]. Especially, the latter characteristic reduces the dead volume of the device, which is essential when dealing with an extremely small sample volume.

The pulsatile and discontinuous drug delivery systems are more advantageous in the academic and clinic applications, which can be used in therapeutics as concentrations of solutions vary with time, or matches body's release of peptides and hormones [1,19]. The device would deliver drugs according to various stimuli such as chemical pH, electric fields, and temperature. By judging the degree of external signal, and the device would release appropriate amounts of drug [20,21].

In this paper, the EHD based drug delivery system is applied mainly for three reasons. Firstly, it doesn't require complex fabrication because there is no moving part in the whole device. Secondly, it can produce mono-disperse droplets in a wide range of sizes using either electrospray or single-droplet generation. Thirdly, instability, poor directionality, and diversity in droplet size can be reduced by coating Teflon on surface of PDMS chip and reducing surface tension of liquid in fabrication process [21,23]. Our work shows that a stable, repeatable and controllable generation of volume at a constant rate has been achieved by the EHD method. Many types of solutions can be delivered through the device such as DI water, acetic acid, methanol, and so on.

## II. Device Concept and Operating Principle

A typical EHD based micro drug delivery device is shown in Fig.1. An electric field would be formed after when we apply voltage on two electrodes. Drug solutions would be under the electric pressure as soon as they arrive at the tip of the red nozzle. Under the pressure, the liquid is elongated and breaks up into single micro-droplets from the tip. Taylor cone, which is very common in hydrodynamic spray processes, is suppressed by controlling the surface wetting property of the PDMS device and the surface tension of the sample liquids.

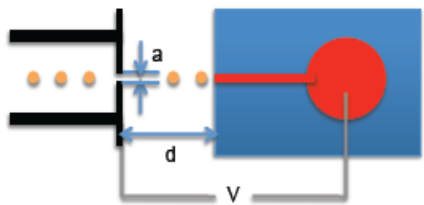

Fig.1 Schematic of EHD drug delivery device

## III. Model Design and Fabrication
1. Device Design and Dimensions
The core of the device made of PDMS consists of a spherical drug reservoir and a cylindrical tube extruded from the reservoir, as shown in Fig.2. A metal tube/needle, as an anode, is inserted into the PDMS core.

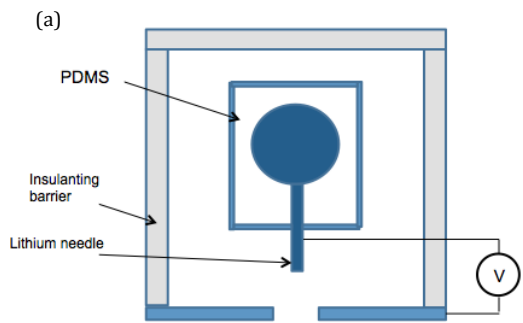

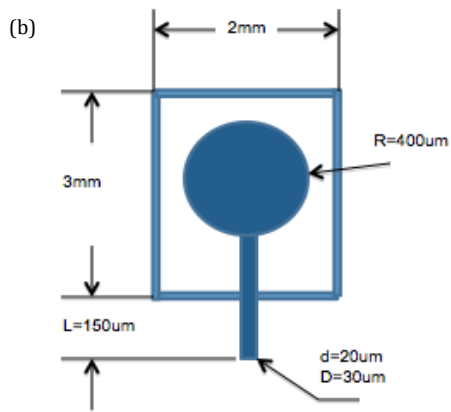

Fig.2 Design and dimensions of EHD based device. (a) PDMS core (blue), external box (grey) and applied voltage (black); (b) dimensions of PDMS core and metal tube.

2. Materials
Electrode: Silicon-based electrode coated with polyimide over an adhesion promoter.
Insulating material: Diamond-like carbon (DLC) film adheres strongly to some plastics providing DLC is biocompatible.

3. Fabrication
The big chunk of PDMS in the middle is supposed to be made up by two half parts. For the reservoir, it is a half sphere, and for the micro-channel, it is a half cylinder. Overall, the fabrication will be complicated. So we choose to use Grey Scale method on SU-8 of E-beam lithography.

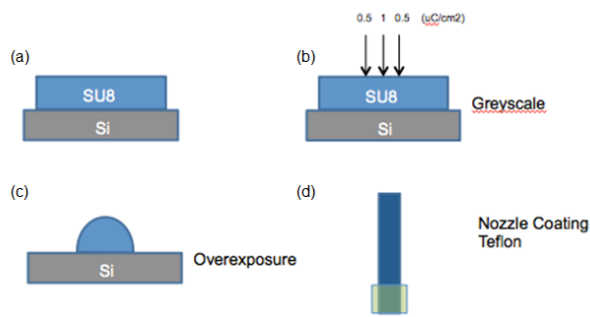

Fig.3 Fabrication process of drug reservoir and Teflon coating at tip of nozzle

Teflon coated on the inner and outer surface of tip of the metal nozzle, as shown in Fig.3d, can improve the stability and directionality of the droplet formation, and further reduce the droplet size by reducing surface tension of liquid.

Fabrication process, as shown in Fig.3, consists of four steps as follows.
(1) Put SU-8 E-beam resist on top of silicon wafer.
(2) Grey Scale to overexpose the middle, and by manipulating the gradient density to get a perfect half cylinder and a half sphere. In Fig.3b, 1 and 0.5 µC/cm², is just a trial number.
(3) After overexposure, one half core is done. Then cast PDMS on top of it.
(4) Coat metal nozzle with Teflon.

## IV. Charged Droplet Neutralization

Droplets generated by EHD method would carry a slight amount of charges. Since nobody has really taken the droplet neutralization effect into account for a matured drug delivery system, and this part is hard to accomplish. Here, we give out three possible solutions for this neutralization problem.
1. Negative Corona Discharge

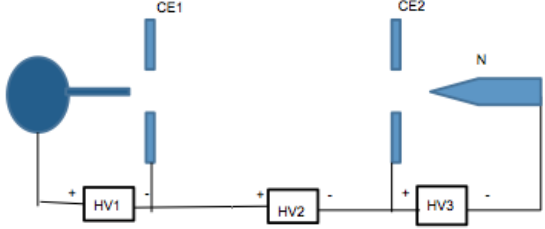

Fig.4 Schematic of a neutralization system

When the charged droplet comes out of CE1, by changing HV3, we can actually change the negative electrons released from tip N, which are used to neutralize the positive charges carried on droplet [25].
2. Negative Spraying of Volatile Liquid
By replacing the N part with a capillary of spraying volatile liquid out, which will bring negative electrons in the liquid, when the evaporation is finished, electrons remain to neutralize the positive charges.
3. Electron Cloud
Electron sources may be classified into (1) electrons produced at the injection region stripping foil; (2) electrons produced by proton losses incident the vacuum chamber at grazing angles; (3) secondary electron emission process; and (4) electrons produced by residual gas ionization [25].
Electrons generated at the wall by proton losses are accelerated and decelerated by the beam potential and hit the opposite wall with a net energy gain, producing secondary electrons and forming a charged cloud.

## V. Calculation of Droplet Size and Formation Time

1. Droplet Size
Droplets are formed at the tip of nozzle, that is, the metal tube inserted into PDMS channel. Figure 5 below shows the formation process of a droplet.

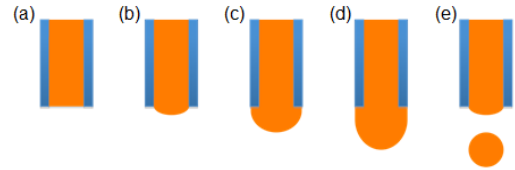

Fig.5 Formation process of a droplet, (a) ~ (e)

Now, assume the geometry of a liquid drop as a spherical cap. It is well known that the volume of a liquid droplet formed at the tip is given by
$$V_{liquid} = V_{cap} = \frac{\pi}{3}h^2(3r - h^2), \quad (1)$$
where, height of a droplet in the larger semisphere, $h = r + \sqrt{r^2 - a^2}$, and radius of nozzle, a.
Then, define the spatial charge density, $\rho_e$, in $C/m^3$. The total charge of a liquid drop is
$$Q = \rho_e \cdot V_{liquid}. \quad (2)$$
In MEMS device, the gravitational effect on a droplet can be neglected. Therefore, there are two groups of forces exerted on a forming droplet, electric force ($F_e$) and surface force ($F_s$), as shown in Fig.6.
Electric force caused by the external electric field, due to Coulomb's law, is given by
$$F_e = E \cdot Q = E\rho_e V_{liquid}. \quad (3)$$
Surface force due to the surface tension between liquid, gas and nozzle, is

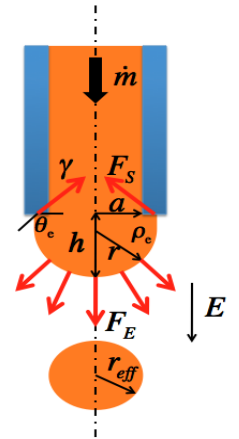

Fig.6 Parameters used in calculation of a droplet formation

given by
$$F_s = \gamma \cdot 2\pi a \cdot \sin(\pi - \theta_c) = 2\pi a \gamma \sin\theta_c, \quad (4)$$
where, contact angle, $\theta_c = \frac{\pi}{2} + \arccos\left(\frac{a}{r}\right)$.

A drug droplet is formed at the tip of nozzle, only when the electric force caused by EHD effect is larger than the surface force on the meniscus of droplet, $F_e > F_s$. The droplet is elongated and breaks up from the nozzle at force equilibrium. The effective radius of a droplet can be calculated as follows.

In equilibrium, $F_e = F_s$,
$$\begin{cases} F_e = \frac{\pi}{3} E \rho_e h^2 (3r - h^2) \\ F_s = 2\pi a \gamma \sin\theta_c \end{cases}, \quad (5)$$
So, volume of a droplet is
$$V_{droplet} = \frac{4}{3}\pi r_{eff}^3 = \frac{2\pi a \gamma \sin\theta_c}{E\rho_e}, \quad (6)$$
and effective radius is
$$r_{eff} = \left(\frac{3a\gamma\sin\theta_c}{2E\rho_e}\right)^{1/3}. \quad (7)$$

So far, we have calculated the effective radius of a formed droplet in equilibrium at static state. Based on our results, the drop size is dependent on nozzle size, drug solution properties (surface tension), Teflon coating (contact angle), spatial charge density, distance between nozzle and ground electrode, and supplied voltage.

2. Droplet Formation Time

Two different time constants are of our interests. $t_{gap}$, the required time between two formed droplets, is given by
$$t_{gap} = \frac{\rho_d V_{droplet}}{\dot{m}} = \frac{2\pi a \gamma \rho_d \sin\theta_c}{E\rho_e \dot{m}}, \quad (8)$$
Which significantly depends on the drug supply rate, $\dot{m}$.

$t_{form}$, the required time to form a single droplet from the tip of nozzle,
$$t_{form} = \frac{V_{droplet}}{\int_0^{2\pi}\int_0^a v(r,\theta)r\,dr\,d\theta} = \frac{V_{droplet}}{\int_0^a v(r)2\pi r\,dr}, \quad (9)$$
where, $v(r, \theta)$ is fluid velocity profile created at the tip of nozzle. We will show how to calculate the velocity profile in the next section.

3. Velocity Profile inside Nozzle

In order to estimate the flow rate inside the electrode induced by an external voltage, we use a well established model called Electro-osmatic flow (EOF), which is the motion of liquid induced by an applied electric field across a charged micro-channel.

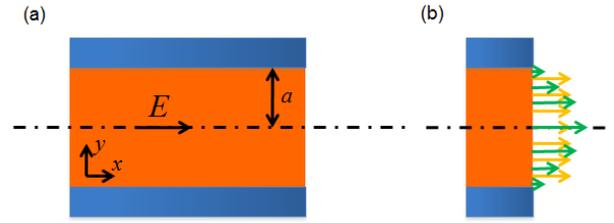

Fig. 7 EOF model. (a) drug flow under constant electric field; (b) two possible velocity profiles at outlet, parabolic (green) and constant (yellow) velocity.

Several assumptions are made for the calculation of velocity profile at tip.
(1) Two-dimensional fully developed flow.
(2) Zero pressure gradient.
(3) Slip velocity at channel wall is $u = \frac{\varepsilon \zeta}{\mu}\vec{E}$, where, $\varepsilon$, drug fluid permittivity, $\mu$, drug fluid viscosity, and $\zeta$, surface electric potential.

The governing equations used in our calculation are shown below,
$$\begin{cases} \nabla \cdot u = 0 \\ \rho u \cdot \nabla u = -\nabla p + \mu \nabla^2 u \end{cases}. \quad (10)$$
Write the continuity and momentum equations in Cartesian form:
$$\begin{cases} \frac{\partial u}{\partial x} + \frac{\partial v}{\partial y} = 0 & (11a) \\ \rho\left(u\frac{\partial u}{\partial x} + v\frac{\partial u}{\partial y}\right) = -\frac{\partial p}{\partial x} + \mu\left(\frac{\partial^2 u}{\partial x^2} + \frac{\partial^2 u}{\partial y^2}\right) & (11b) \\ \rho\left(u\frac{\partial v}{\partial x} + v\frac{\partial v}{\partial y}\right) = -\frac{\partial p}{\partial y} + \mu\left(\frac{\partial^2 v}{\partial x^2} + \frac{\partial^2 v}{\partial y^2}\right) & (11c) \end{cases}$$

Simplify the above equations, we can get
$$v = 0 \to \frac{\partial u}{\partial x} = 0 \to u = u(y). \quad (12)$$
Using the assumption of zero pressure gradient, Eq.(11c) vanishes and Eq.(11b) can be simplified as:
$$0 = \mu \frac{\partial^2 u}{\partial y^2}. \quad (13)$$
A general solution for Eq.(13) is in a form of
$$u = C_1 y + C_2. \quad (14)$$
Apply the boundary conditions, we get:
$$\begin{cases} u(-a) = -\frac{\varepsilon\zeta}{\mu}E = -C_1 a + C_2 \\ u(a) = -\frac{\varepsilon\zeta}{\mu}E = C_1 a + C_2 \end{cases}. \quad (15)$$
Therefore, $C_1 = 0$ and $C_2 = -\frac{\varepsilon\zeta}{\mu}E$.

The fluid velocity is given by

$$u(y) = -\frac{\varepsilon\zeta}{\mu}E. \quad (16)$$

It would be much easier to calculate the formation time as velocity is constant, since no integration is needed

$$t_{form} = \frac{V}{uA} = \frac{V_{droplet}}{\frac{\varepsilon\zeta}{\mu}E\pi a^2} = \frac{2\gamma\mu\sin\theta_c}{aE^2\rho_e\varepsilon\zeta}. \quad (17)$$

Eq.(17) suggests the formation time of a droplet is a function of surface tension coefficient, viscosity, contact angle, nozzle size, electric field strength, permittivity, etc.

**VI. COMSOL Simulation and Results**

Here, nozzle size $a = 10$ μm, applied voltage $V = 1500$ V, gap $d = 100$ μm. So, $E = 1.5 \times 10^7$ V/m. The liquid used in the simulation is water, the common solution to most drug solutions. Its surface tension $\gamma = 72 \times 10^{-3}$ N/m, and after Teflon coating, the contact angle increases, $\theta_c = 110°$. Assume the spatial charge density of a droplet $\rho_e = 1 \times 10^{-4}$ C/m$^3$. Simulations of EHD based device are shown in Fig.8.

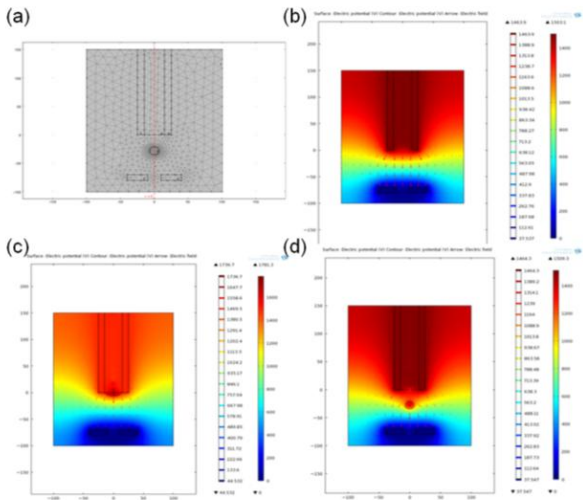

Fig.8 COMSOL simulation of electric field distribution between a metal needle and ground electrode. (a) numerical analysis domain. (b) initial setup. (c) beginning of droplet formation process. (d) a liquid droplet formed by EHD.

We also simulated the neutralization process, which is necessary for a newly formed droplet before ejecting into the body. Highly charged droplets formed by EHD method and the neutralization region with opposite charges are shown in Fig.9.

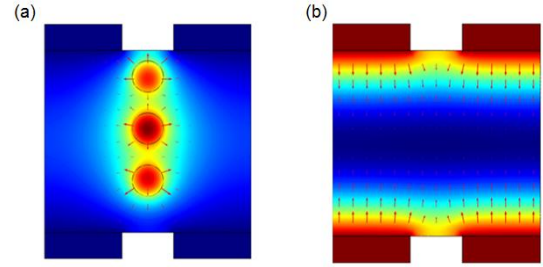

Fig.9 Simulation of the neutralization region. (a) high charged droplets without neutralization. (b) neutralization region with opposite charges.

Different ratios of the spatial charge density on droplets and neutralization region, affect the neutralization effect, as shown in Fig.10. The ratio ranges from 4000:1 to 1:2. Neutralization works well when the ratio is around 200:1, at which a formed droplet is neutralized completely. Otherwise, the charges generated by EHD on droplet are either incompletely neutralized or changed to the opposite because of the higher charge density in neutralization region.

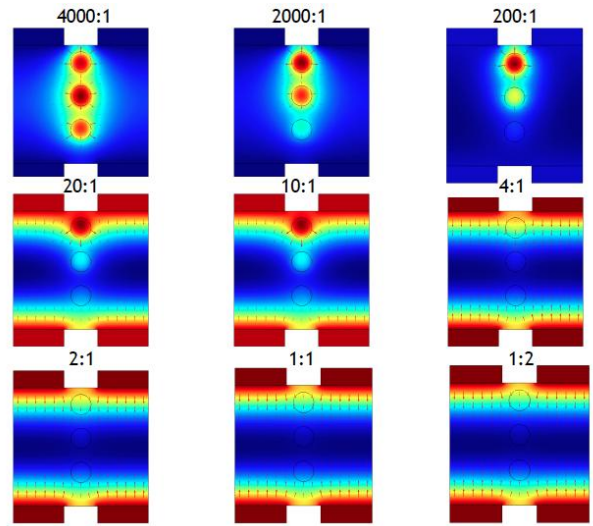

Fig.10 Effect of neutralization due to ratios of spatial charge density on droplets and neutralization region

**VII. Summary**

Our EHD pulsatile drug delivery system can deliver a single droplet as small as 2.83 nL in 8.5 ms with a total device of $2 \times 2 \times 3$ mm$^3$, and a working voltage of 1500V. The droplet generation is repeatable and stable.

Table 1 Dimensions and system data used in calculation

| Dimensions of device (μm) | | | |
|---|---|---|---|
| O.D. | I.D. | $L_{needle}$ | $R_{res}$ |
| 30 | 20 | 150 | 400 |
| $W_{PDMS}$ | $L_{PDMS}$ | a | d |
| 2 | 3 | 10 | 100 |
| External voltage (V) | | 1500 | |
| Drug supply rate (kg/min) | | 0.02 | |
| Spatial charge density (C/m$^3$) | | $1 \times 10^{-4}$ | |
| Surface electric potential (V) | | 0.1 | |

Except for DI water we used in calculation and simulation, many other solutions could be used in our design. Properties of various solutions [21] are shown in table 2 below.

Table 2 Properties of commonly used drug solutions

| | Density $\rho$ (g/cm$^3$) | Surface tension $\gamma$ (mN/m) | Permittivity $\epsilon/\epsilon_o$ | Viscosity $\mu$ (mPa·s) |
|---|---|---|---|---|
| acetic acid | 1.001 | 70.37 | 78.34 | 1.003 |
| formic acid | 1.002 | 71.19 | 78.52 | 1.004 |
| methanol | 0.998 | 67.16 | 78.47 | 1.011 |
| ethanol | 0.995 | 67.16 | 78.45 | 1.005 |
| DI water | 1.000 | 73.05 | 78.54 | 1.002 |

## VIII. Conclusion

Electrohydrodynamic generation, a commonly used method in BioMEMS, plays a significant role in the pulsatile drug delivery system for a decade.

In this paper, an EHD based drug delivery system is well designed for the stable, repeatable and controllable generation of a nano/micro droplet at a constant rate. Theoretically, we derive the expressions for the size and the formation time of a droplet generated by EHD method, while taking into account the mass supply rate and properties of drug, gap between two electrodes, nozzle size, and charged droplet neutralization.

Even with an inserted metal tube or needle to enhance the efficiency and stability of the droplet formation, its fabrication and critical alignment remain challenging.